\documentclass[conference]{IEEEtran}
\IEEEoverridecommandlockouts
\usepackage{cite}
\usepackage{amsmath,amssymb,amsfonts}
\usepackage{algorithmic}
\usepackage{graphicx}
\usepackage{textcomp}
\usepackage{xcolor}
\def\BibTeX{{\rm B\kern-.05em{\sc i\kern-.025em b}\kern-.08em
    T\kern-.1667em\lower.7ex\hbox{E}\kern-.125emX}}

\begin{document}

\title{COVID-Net MLSys: Designing COVID-Net for the Clinical Workflow}


\author{Audrey G. Chung$^{2,3}$, Maya Pavlova$^{1}$, Hayden Gunraj$^{1}$, Naomi Terhljan$^{1}$, Alexander MacLean$^{1,2}$,\\ Hossein Aboutalebi$^{1,2}$, Siddharth Surana$^{1}$, Andy Zhao$^{1}$, Saad Abbasi$^{1}$, and Alexander Wong$^{1,2,3}$\\
$^1$Vision and Image Processing Research Group, University of Waterloo\\
$^2$Waterloo Artificial Intelligence Institute, University of Waterloo\\
$^3$DarwinAI Corp.\\
}




\maketitle

\begin{abstract}
As the COVID-19 pandemic continues to devastate globally, one promising field of research is machine learning-driven computer vision to streamline various parts of the COVID-19 clinical workflow. These machine learning methods are typically stand-alone models designed without consideration for the integration necessary for real-world application workflows. In this study, we take a machine learning and systems (MLSys) perspective to design a system for COVID-19 patient screening with the clinical workflow in mind. The COVID-Net system is comprised of the continuously evolving COVIDx dataset, COVID-Net deep neural network for COVID-19 patient detection, and COVID-Net S deep neural networks for disease severity scoring for COVID-19 positive patient cases. The deep neural networks within the COVID-Net system possess state-of-the-art performance, and are designed to be integrated within a user interface (UI) for clinical decision support with automatic report generation to assist clinicians in their treatment decisions.
\end{abstract}

\section{Introduction}
The 2019 coronavirus pandemic (COVID-19), caused by severe acute respiratory syndrome coronavirus 2 (SARS-CoV-2), continues to have a devastating global impact with far-reaching social and economic effects as shown by the World Health Organization~\cite{WHO2020}. In particular, the COVID-19 pandemic has placed a tremendous burden on healthcare systems around the world. The systems are struggling to keep up with providing care and treatment to patients due to limited health workers and clinical resources such as mechanical ventilators, oxygen, personal protective equipment, and other medical supplies.

A critical step in the COVID-19 clinical workflow for patient triaging and care planning~\cite{Emanuel2020,Li2020,Tyrrell2021} is effective screening methods.  The current standard method for COVID-19 screening is transcriptase-polymerase chain reaction (RT-PCR) testing~\cite{Wang2020_RTPCR}, where the ribonucleic acid (RNA) of the SARS-CoV-2 virus is detected based on an upper respiratory tract sputum sample.  While the specificity of RT-PCR is high, a number of recent studies have found that the sensitivity of RT-PCR tests can be relatively low and fluctuate greatly depending on how or when the specimen was collected~\cite{Fang2020,Yang2020, Li2020_RTPCR, Ai2020}.

Chest radiography screenings have seen significant growing interest and use in healthcare systems worldwide as a complimentary screening method to RT-PCR~\cite{Wong2020_CXR,Warren2018,Toussie2020,Huang2020,Guan2020}. Chest radiography holds a number of important benefits as a screening tool for COVID-19. First of all, chest radiography equipment is one of the most widely available and accessible medical imaging modalities in many healthcare facilities worldwide. Second, chest radiography equipment is relatively fast to decontaminate compared to other medical imaging equipment, and the widespread availability of portable systems~\cite{Jacobi} enable screening to be conducted in isolation rooms to greatly lower the chances of transmission~\cite{RSNA}. Third, chest radiography is routinely used to assess respiratory complaint~\cite{BSTI}, which is one of the key symptoms of COVID-19, and therefore can be used in parallel to time-consuming RT-PCR tests. Finally, chest radiography allows for the assessment of the severity of the condition of a COVID-19 positive patient, which cannot be done with RT-PCR tests.

\begin{figure*}[t]
\begin{center}
\includegraphics[width=\linewidth]{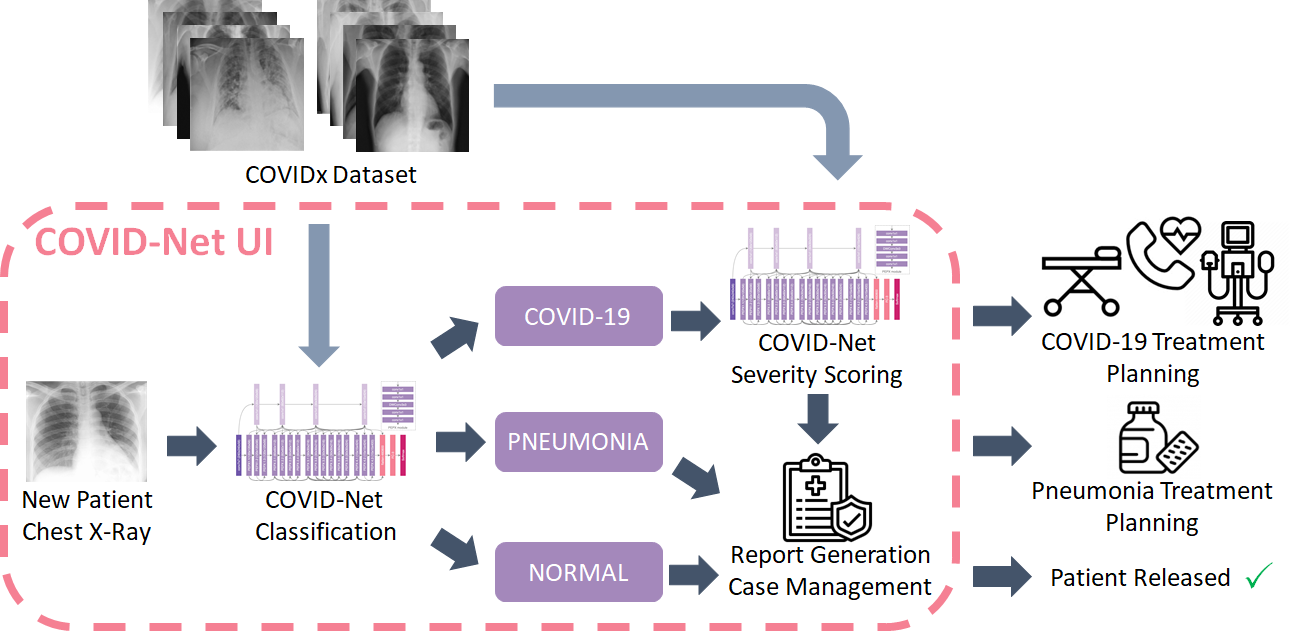}
\end{center}
\caption{Overview of the clinician-centric COVID-Net system. The COVIDx dataset is used to train the COVID-Net deep neural networks for COVID-19 detection and severity scoring outside of the clinical workflow. The main clinical workflow (i.e. detection of COVID-19 patient cases, severity scoring of COVID-19 positive patient cases, and report generation for new patient chest X-rays) is wrapped in our COVID-Net UI, as indicated by the pink dashed box. For each patient, the medical professional can leverage the outputs of the COVID-Net system and plan the corresponding treatment.}
\label{fig:algdesign}
\end{figure*}

Despite the numerous benefits of chest radiography for COVID-19 screening, one of the biggest challenges is the limited number of expert radiologists who can be made available to interpret the imaging data to perform screening and severity assessment.  As such, the availability of computer-aided clinical decision support systems for aiding radiologists to accelerate the speed of interpretation, as well as for clinicians and front-line healthcare workers to directly interpret the imaging data, can greatly assist the COVID-19 clinical workflow to better care for the large number of patients and better manage the current COVID-19 pandemic.

Motivated to assist in the fight against COVID-19 and in light of the tremendous benefits and utility of chest radiography for COVID-19 screening and treatment planning, we launched the COVID-Net Open Source Initiative~\footnote{COVID-Net Open Source Initiative: http://covid-net.ml/}~\cite{Wang2020,Wong2020,Gunraj2020,Gunraj2021} for accelerating the advancement and adoption of deep learning for tackling the COVID-19 pandemic. While the initiative has been successful and leveraged globally, we have thus far presented disparate contributions to directly assist radiologists with the challenges in differentiating between COVID-19 positive infections, non-COVID pneumonia infections, and normal chest images.

In this study, we present the COVID-Net system, a machine learning system for COVID-19 classification and severity scoring using chest X-rays (CXRs), that is developed with the clinical workflow in mind. Taking a machine learning and systems (MLSys)~\cite{Ratner2019} approach, the COVID-Net system integrates state-of-the-art predictive models within our user interface (UI) for clinical decision support, taking computer vision from theory to tangible reality.

\section{The COVID-Net System}
Figure~\ref{fig:algdesign} presents an overview of our clinician-centric COVID-Net system. The COVID-Net system comprises of the COVIDx dataset of chest X-ray images (CXRs), the COVID-Net deep neural network for detecting COVID-19 patient cases, and the COVID-Net S deep neural networks for scoring the disease severity of COVID-19 positive patient cases. The COVIDx dataset is used to train the COVID-Net and COVID-Net S deep neural networks outside of the clinical workflow, and the main clinical workflow is wrapped in a user interface (UI) clinical decision support system we designed with automatic report generation to assist clinicians in their treatment decisions. The individual components of the COVID-Net system are described below.

\begin{figure*}[t]
\begin{center}
     \includegraphics[width=0.95\linewidth]{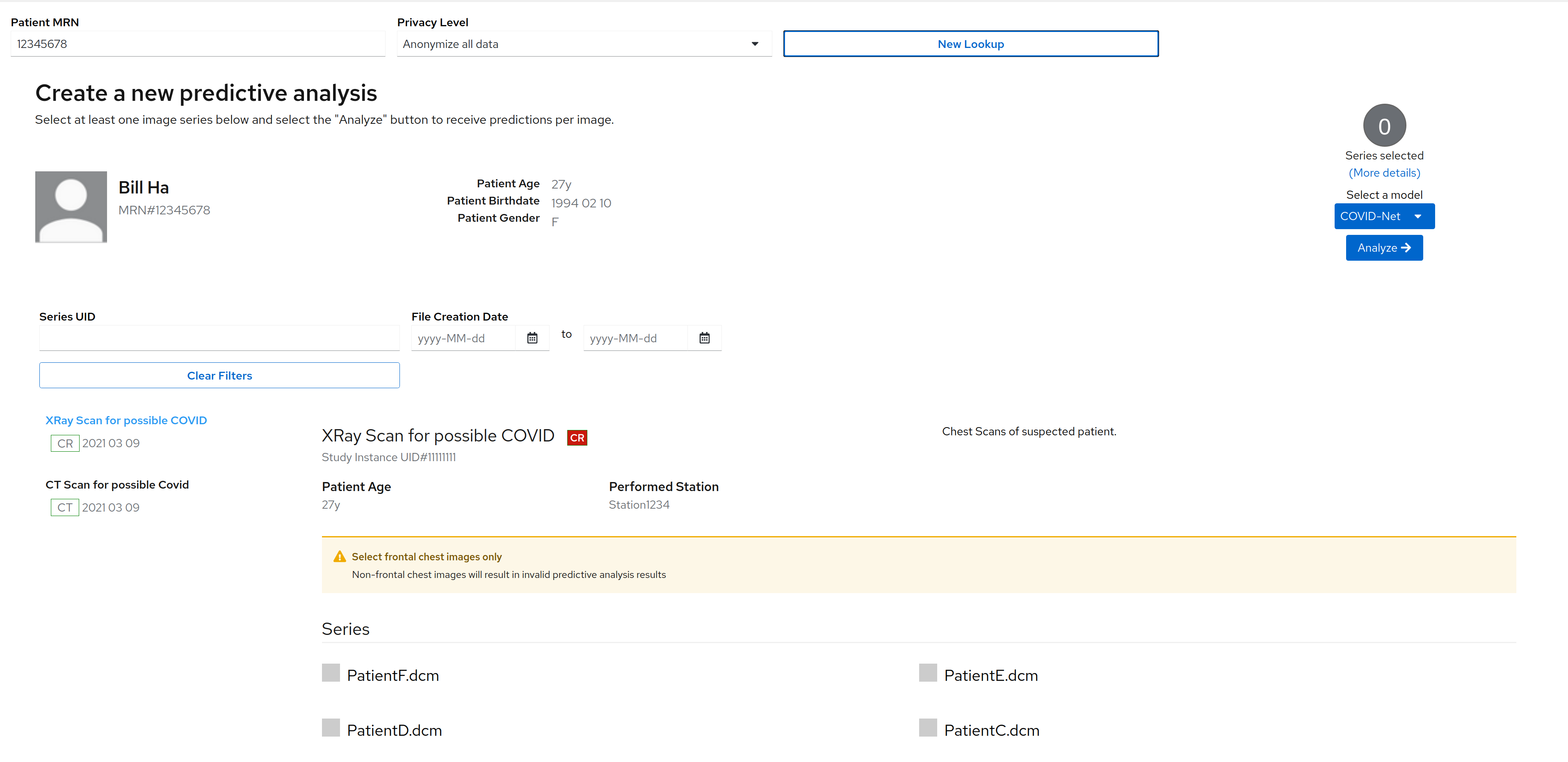} \\
     \includegraphics[width=0.95\linewidth]{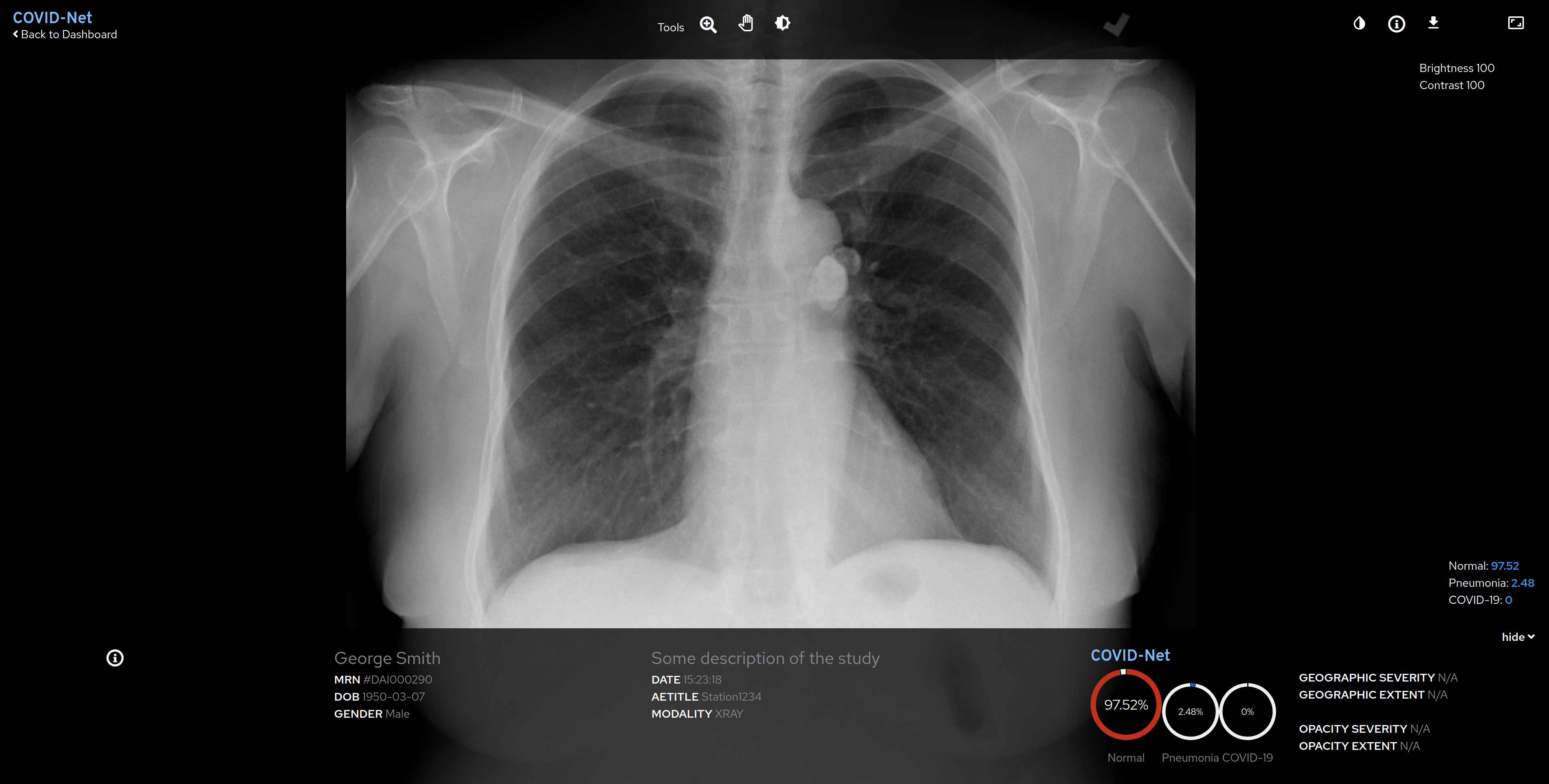}
\end{center}
\caption{Sample screens of the COVID-Net user interface (UI). The COVID-Net user interface (UI) was designed and developed so that medical professionals and researchers can query, view, and analyze patient scans in a more streamlined way.}
\label{fig:GUI}
\vspace{-10pt}
\end{figure*}

\subsection{The COVIDx dataset}

In order to build the different COVID-Net deep neural networks for both detection and severity scoring of COVID-19 patient cases, the first step is to construct a large, diverse corpus of CXR images comprising patient cohorts from around the world to improve generalization of the built deep neural networks.  As such, the first core component of the COVID-Net system is COVIDx, a dataset of CXR images curated specifically for building computer-aided clinical decision support algorithms for the COVID-19 clinical workflow.  More specifically, in its current state, the COVIDx dataset~\cite{Wang2020} used to build the COVID-Net system  is composed of 16,560 CXR images from 15,528 patients from at least 51 countries, making it one of the largest, most diverse COVID-19 CXR datasets in open access form. The COVIDx dataset continues to grow as more data is curated from different clinical data sources from around the world.

\subsection{COVID-Net: Deep neural network for COVID-19 patient case detection}

The second core component in the COVID-Net system is the COVID-Net deep neural network~\cite{Wang2020}, which is responsible for COVID-19 patient case detection.  More specifically, the COVID-Net deep neural network possesses a highly customized deep convolutional neural network architecture with macroarchitecture and microarchitecture designs that are tailored using a machine-driven design exploration strategy~\cite{gensynth}. The model differentiates between COVID-19 positive patients, patients with non-COVID pneumonia infections, and normal control patients. COVID-Net has an accuracy of 93.3\%, with a COVID-19 sensitivity of 91.0\% and a positive predictive value of 98.9\%. Furthermore, an explainability-driven performance analysis was conducted on COVID-Net via a qualitative explainability method~\cite{gsinquire} to ensure that the critical factors leveraged during the decision-making process are clinically relevant (e.g., ground-glass opacities, bilateral abnormalities) and not erroneous visual cues (e.g., embedded text, imaging artifacts, etc.).

\subsection{COVID-Net S: Deep neural networks for disease severity scoring of COVID-19 positive patients}

The third core component in the COVID-Net system is the pair of COVID-Net S deep neural networks~\cite{Wong2020}, which are responsible for scoring the lung disease severity of a COVID-19 positive patient.  More specifically, the pair of COVID-Net S deep neural networks possess highly customized deep convolutional neural network architectures that are tailored for quantifying the lung severity based on two key metrics (one network for each metric): i) \textbf{geographic extent}: aggregate extent of lung involvement by ground glass opacity or consolidation for both lungs (scale of 0-8, with 0 being no involvement and 8 being $>$75\% involvement in both lungs), and ii) \textbf{opacity extent}: aggregate degree of opacity for both lungs (scale of 0-8, with 0 being no opacity and 8 being complete white-out in both lungs).

A stratified Monte Carlo cross-validation
testing strategy was employed involving 100 experiments to evaluate severity scoring performance, and it was found that the COVID-Net S networks yielded R$^2$ of $0.664 \pm 0.032$ and $0.635 \pm 0.044$ between predicted scores and radiologist scores for geographic extent and opacity extent, respectively. The best performing COVID-Net S networks achieved R$^2$ of 0.739 and 0.741 between predicted scores and radiologist scores for geographic extent and opacity extent, respectively, and these best performing networks are utilized in the COVID-Net system.

\subsection{COVID-Net UI: Report generation and case management}

The fourth and final core component of the COVID-Net system is the COVID-Net user interface (UI), a report generation and case management platform  designed and developed so that medical professionals and researchers can query, view, and analyze patient scans in a more streamlined way (see Figure~\ref{fig:GUI}). A key functionality of the UI is automatic report generation to assist clinicians in their treatment decisions. The COVID-Net UI, with integration of the COVID-Net and COVID-Net S deep neural networks, is currently in the process of designing a user testing study with medical professionals at clinical institutions. In particular, we are gathering feedback from users to assess if the COVID-Net UI works in ways that are intuitive and would fit into a clinical workflow.

\section{Conclusion}
We presented the COVID-Net system, a machine learning system for COVID-19 classification and severity scoring developed with the clinical workflow in mind. The systems comprises of the COVIDx dataset, the COVID-Net deep neural network for detecting COVID-19 patient cases, the COVID-Net S deep neural networks for lung severity scoring, and the COVID-Net UI for automatic report generation and case management. The COVID-Net system integrates state-of-the-art deep neural networks for different tasks in the clinical decision support workflow into a user interface (UI) we designed with automatic report generation functionality to assist clinicians in their treatment decisions. We are currently designing a user testing study with medical professionals to assess if the COVID-Net UI is intuitive and fits into a clinical workflow.

Future works include incorporating patient electronic health record (EHR) information to better integrate with the clinical workflow, extending the COVID-Net system to other imaging modalities (e.g., computed tomography and ultrasound), and implementing the COVID-Net system to support multiple classification and severity scoring models.

\bibliographystyle{ieee_fullname}
\bibliography{COVIDNet_Pipeline}

\end{document}